# A bulk acoustic resonator with vertical electrodes for wideband filters

Silvan Stettler[1,*], Edgar Navarro-Gesse[2], Carlos Collado[2], Jordi Mateu[2], Luis G. Villanueva[1,†]

[1] Advanced Nanoelectromechanical Systems Laboratory, École Polytechnique Fédérale de Lausanne (EPFL), 1015 Lausanne, Switzerland

[2] Signal Theory and Communications Department, Universitat Politècnica de Catalunya (UPC), Castelldefels, Spain

[*] Corresponding author; silvan.stettler@epfl.ch
[†] guillermo.villanueva@epfl.ch

**Abstract**

Radiofrequency (RF) front ends for current and next generation (5G and 6G) wireless communication demand acoustic filters that combine wide bandwidth, high power capability, and thermal stability. Existing surface and bulk acoustic wave (SAW and BAW) technologies face inherent trade-offs between electromechanical coupling, lithographic tunability, and robustness. Here we introduce the bulk acoustic resonator with vertical electrodes (VBAR), a device that combines the advantages of suspended and solidly mounted resonators. VBARs use lithium niobate (LiNbO$_3$) ridges with sidewall electrodes to excite a shear-horizontal bulk acoustic resonance, providing frequency control through lithography in a configuration that is mechanically anchored to the substrate. Fabricated VBARs exhibit electromechanical coupling coefficients exceeding 30% in the 2–4 GHz range, enabling ladder filters with fractional bandwidths of nearly 20%. While further optimization is necessary to minimize losses, the VBAR concept offers an alternative route toward wideband and robust RF filters for next-generation wireless systems.

**Introduction**

Currently, the radiofrequency (RF) filter market is dominated by surface acoustic wave (SAW) and bulk acoustic wave (BAW) technologies, which have been continuously improved to comply with the evolving mobile communication standards. The roll-out of the fifth generation (5G) standard has created a renewed demand for microacoustic filters capable of delivering higher frequencies, enhanced fractional bandwidths (FBW), low temperature coefficients of frequency (TCF), and increased power handling[1]. Some bands in the sub-6GHz spectrum (5G Frequency Range 1, 5G-FR1) demand not only higher frequency than previous LTE bands, but also much larger fractional bandwidths (FBW, up to 24% for n77). In addition, GPP3 Release 17 introduced power class 1.5 which specifies increased output powers of 29 dBm for some of those bands (i.e. n77, n78)[2]. These specifications have been driving the progress of SAW and BAW resonator technology in recent years, each offering distinct advantages in certain aspects while exhibiting limitations in others. For common filter topologies such as the ladder filter, resonance frequency ($f_r$), electromechanical coupling coefficient ($k_t^2$) and quality factor ($Q$) of the individual resonator building block relate to the center frequency ($f_c$), fractional bandwidth (FBW) and insertion loss (IL) of the filter[3].

For SAW resonators, $f_r$ can be scaled lithographically by varying the pitch of the interdigital transducer (IDT) electrodes. This dependence makes SAW technology extremely versatile, as it enables the design of multiple filtering components with different $f_c$ on the same die (e.g. for multiplexers) by changing the IDT layout. In addition, the most widely used SAW substrates are strong piezoelectrics such as lithium niobate (LiNbO$_3$) or lithium tantalate (LiTaO$_3$), which can provide large $k_t^2$ (> 20%) resulting in FBWs of over 15% in a ladder filter[4]. Unfortunately, the use of conventional SAW technology based on bulk piezoelectric substrates for the frequencies > 2.5 GHz (necessary for 4G and 5G) is complicated because of the losses due to bulk wave radiation and low acoustic phase velocities ($v_p$). In this regard, one of the key innovations in recent years has been the introduction of thin film piezoelectric on insulator (POI) substrates[5–7]. These substrates consist of a thin film of monocrystalline LiNbO$_3$/LiTaO$_3$ bonded to a high velocity substrate such as high-resistivity Si[8–10], quartz[11,12], sapphire[13] or silicon carbide (SiC)[14–18] and have led to significant improvements of the quality factor ($Q$), $k_t^2$, and resonance frequencies > 2.5 GHz. Even though the lithographic resolution is still a limiting factor, POI substrates have enabled SAW-based filters with low IL above 5 GHz[19].

In contrast to SAW, the frequency in BAW resonators such as free-standing bulk acoustic resonators (FBAR) or solidly mounted BAW resonators (BAW-SMR) is not lithography dependent but controlled by the thickness of the piezoelectric film instead. Commercial BAW products rely on AlN and scandium-doped AlN (AlScN) which can be deposited in sub-micron thickness layers to reach frequencies up to 8 GHz and above[20–22]. The intrinsic piezoelectric coupling factor of AlScN limits the $k_t^2$ of such BAW resonators to around 20%, thereby restricting wide-bandwidth potential. It has been shown that LiNbO$_3$-based BAW resonators can achieve much higher $k_t^2$ (> 30%) than all previously mentioned technologies. However, integrating monocrystalline LiNbO$_3$ films into an FBAR or BAW-SMR structure requiring metal layers below the film poses technological challenges[23–26]. Laterally excited bulk acoustic resonators (XBARs) have emerged as a promising solution in that regard[27–31]. XBARs benefit from the high $k_t^2$ of the shear BAW in LiNbO$_3$ and scalability to high frequency, but do not require a metal layer below the suspended piezoelectric film[29,32,33]. Despite its advantages, the XBAR architecture, employing a suspended LiNbO$_3$ film, is not conducive for sustaining high power levels[34]. In comparison, solidly mounted technologies, such as BAW-SMR and POI-based SAW, are inherently more robust and heat can be dissipated away from the surface owing to the high thermal conductivity of commonly used substrates (Si ~ 140 W/(K·m), SiC > 400 W/(K·m)). In a suspended film, heat generated under high power can only be dissipated laterally through the film (or through the narrow electrodes) and to exacerbate the issue, LiNbO$_3$ is a poor thermal conductor (< 10 W/(K·m)).

In this paper, we explore an alternative resonator structure[35] which can achieve similar levels of $k_t^2$ as LiNbO$_3$-based BAW resonators, but with a solidly mounted configuration and the capability to set the resonance frequency by lithography (like SAW resonators). Recently, it was shown that vertical Si fins (attached to the substrate) with piezoelectric transducers on the sidewalls can support lithographically defined BAW resonances[36], albeit with low $k_t^2$. Here, we introduce the bulk acoustic resonator with vertical electrodes (VBAR), which consists of LiNbO$_3$ piezoelectric ridges with metal electrodes deposited on the sidewalls. We experimentally demonstrate that (lateral shear) BAW resonances with $k_t^2$ over 30% in the 2-4 GHz range can be excited in such a structure and we show that ladder filters using VBARs as building blocks can attain FBWs close to 20%. Furthermore, we investigate the TCF and behavior under high power (including non-linearity) of the VBARs. Despite remaining technological challenges, we show that VBARs can sustain RF powers up to 30 dBm.

## Results

**Vertical electrode bulk acoustic resonator (VBAR) concept**

Fig. 1a depicts a schematic illustration of the resonator architecture we propose. The resonating elements consist of an array of piezoelectric ridges that are attached to a silicon substrate via a narrow Si pedestal. A layer of aluminum deposited on the vertical sidewalls of the ridges forms the electrodes to drive the resonator. To address the temperature stability of the resonator, a layer of SiO$_2$ can optionally be integrated on top of the ridge.

Instead of confining the acoustic wave between the horizontal planar surfaces of a suspended membrane

(i.e. FBARs, XBARs), the resonating mode in VBARs is a standing wave confined by the vertical sidewalls of the ridge. Such a configuration presents several differences over BAW resonators utilizing thickness-defined BAWs. The resonance frequency for VBARs is set by the lithographically controllable width of the ridges ($w_{ridge}$) instead of the film thickness. In addition to the planar footprint (i.e. the number of parallel-connected ridges), the static capacitance of a VBAR can be scaled by increasing the thicknesses of the $LiNbO_3$ and $SiO_2$ layers. In contrast to resonators employing suspended films, anchoring the ridges to the substrate provides a path to dissipate heat generated by losses and increased mechanical stability, which may improve power handling capacity.

Since the driving electric field is applied in-plane, we use YX36°-cut $LiNbO_3$ as a piezoelectric material, which possesses strong electromechanical coupling coefficients for shear acoustic waves for excitation with in-plane electric fields[37]. In Fig. 1b, we show the mode shape of the main resonant mode, which is a shear horizontal bulk acoustic wave (SH-BAW). The SH-BAW is characterized by displacement that is out-of-plane ($u_y$, along y in Fig. 1b) and a stress field with a primarily shear horizontal component ($\sigma_{xy}$). The half wavelength ($\lambda/2$) of the SH-BAW roughly corresponds to the total width of the acoustic cavity consisting of the $LiNbO_3$ ridge and the electrodes. The width of the $LiNbO_3$ ridge represents the critical lithographic dimension. The geometry of the Si pedestal, serving as an anchor for the ridges, plays a key role in achieving optimal acoustic performance. Solidly mounted resonators such as BAW-SMR or POI SAW typically use a Bragg reflector stack[38,39] or operate below the cut-off frequency of the BAWs in the substrate[40] to confine the acoustic energy in the piezoelectric layer. Instead, we reduce the leakage of acoustic energy to the substrate by creating a constriction in the anchor below the $LiNbO_3$ ridge. In this particular and unique case of shear vibrations where the displacement of the resonant mode is primarily parallel to the boundaries of the resonator, a constriction can inhibit BAW transmission to the substrate. We note that this approach does not require substrates with high acoustic velocities such as SiC or diamond and can be implemented with cheaper Si substrates. The mode shape in Fig. 1b shows that the displacement at resonance is well confined in the acoustic cavity. In the Supplementary Information, Section 1, we further discuss the rationale this structure and illustrate the influence of the anchor geometry on $Q$, $k_t^2$, and $f_r$. Indeed, in addition to minimized acoustic losses, a narrow anchor allows the LiNbO$_3$ elements to deform in an unconstrained fashion, thereby maximizing $k_t^2$. Since narrow anchors isolate the resonating elements from the substrate, the individual VBAR ridges in the array are acoustically independent of each other. Therefore, the pitch ($p$) of the array does not influence the resonance frequency and can be chosen arbitrarily large to facilitate fabrication.

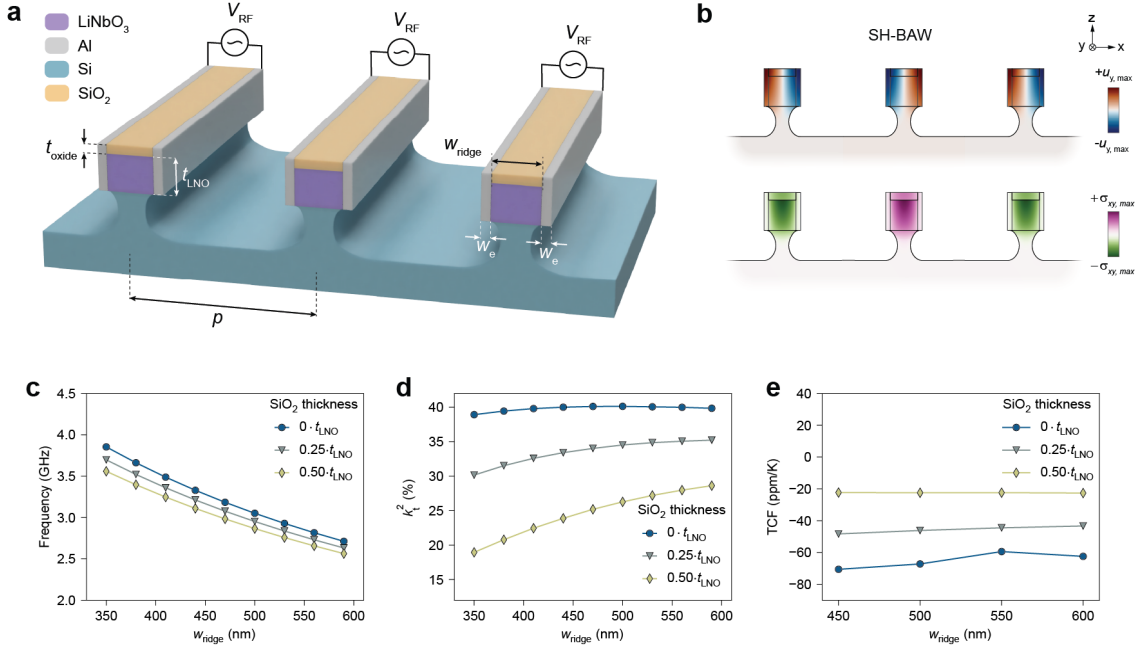

**Figure 1. Vertical electrode bulk acoustic resonator (VBAR) working principle. a** Schematic of a VBAR composed of multiple parallel-connected ridges of LiNbO$_3$ and SiO$_2$ with vertical Al electrodes on the ridge sidewalls. **b** Mode shape of the shear horizontal bulk acoustic wave (SH-BAW) resonance. $u_y$ is the out-of-plane displacement (along y) and $\sigma_{xy}$ is the shear horizontal component of the stress tensor. **c-e** Simulated **c** resonance frequency, **d** electromechanical coupling ($k_t^2$) and **e** first-order temperature coefficient of the resonance frequency of the SH-BAW mode for varying relative thicknesses of the SiO$_2$ layer ($t_{oxide}$). The electrode width ($w_e$) is 100 nm and the LiNbO$_3$ thickness ($t_{LNO}$) is 400 nm.

In Fig. 1c, we show that the resonance frequency can be tuned by adjusting $w_{ridge}$. Assuming Al electrodes with a width of 100 nm on either side, the simulations predict that $f_r$ can be varied from 2.5 to 4 GHz for values of $w_{ridge}$ between 350 to 600 nm. If a layer of SiO$_2$ is added on top of the LiNbO$_3$ for temperature compensation, the phase velocity of the acoustic cavity is lowered, thereby shifting the resonance to slightly lower frequencies. Fig. 1d shows the simulated $k_t^2$ for the same range of $w_{ridge}$ and $t_{oxide}$. Due to the strong piezoelectric coefficient in YX36°-cut LiNbO$_3$, our simulations predict $k_t^2$ close to 40% without a SiO$_2$ layer. If the latter is included, we observe a $k_t^2$ decrease that is approximately proportional to the relative thickness of SiO$_2$. On the other hand, in Fig. 1e we show that the addition of SiO$_2$ can significantly improve the TCF of the SH-BAW mode. Al, Si (at low doping concentrations) and especially LiNbO$_3$ have relatively large negative temperature coefficients of stiffness, which results in a TCF of roughly -70 ppm/K or worse for the uncompensated SH-BAW mode. With its positive temperature coefficient, the SiO$_2$ compensates for the negative temperature coefficient of the other materials involved and we simulate that the TCF can be improved to around -30 ppm/K.

Since the main SH-BAW mode propagates across the width of the ridge, the absolute thickness of the LiNbO$_3$ ($t_{LNO}$) is not the primary factor that determines $f_r$ and $k_t^2$. However, depending on the chosen thickness, thickness-dependent spurious modes might appear close to the main SH-BAW mode. Further simulations that elucidate the impact of the layer thicknesses are given in the Supplementary Information, Section S2.

**VBAR implementation and characterization**

The experimental implementation of VBARs poses some unique challenges that are not commonly encountered for more standard BAW or SAW resonator architectures. First, to accurately control the resonance frequency, the width of the ridges forming the resonating body must be uniform across the thickness of the structure. Like for SAW devices, this requires a well-controlled lithography process, but also a method to pattern the constituent layers with vertical sidewalls. Second, to minimize the leakage of acoustic energy to the substrate, a narrow anchor below the ridge is required. Removing Si from underneath a device is common in fabrication processes for acoustic resonators or MEMS in general and can be achieved with several different methods (such as XeF$_2$ gas, SF$_6$ plasma, potassium hydroxide). For VBARs, the undercut needs to be controlled with sub-micron precision to attain less than half of $w_{ridge}$ to avoid excessive substrate leakage but not much less, otherwise the ridges may be released entirely. Moreover, the undercut needs to be the same for all the individual ridges that compose a VBAR to ensure that they resonate at the same frequency. This process is difficult due to the high etch rate and non-uniformity of many of the release processes (i.e. XeF$_2$). We developed a fabrication process flow for VBARs that addresses those challenges, and we give a detailed description in the Methods.

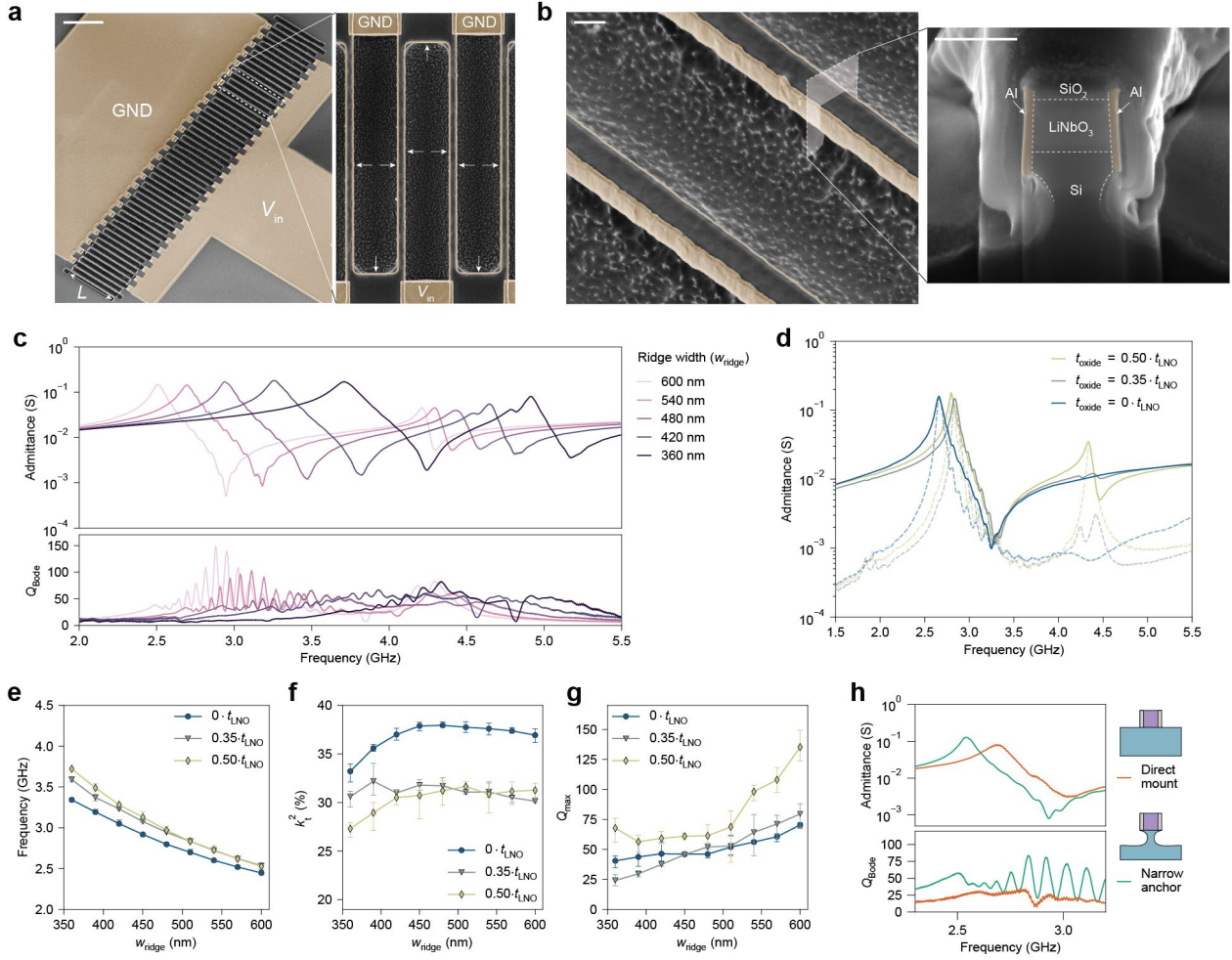

**Figure 2. VBAR implementation and characterization. a** False-color SEM micrographs of a fabricated one-port VBAR. Features composed of aluminum are artificially shaded in dark yellow. The annotations indicate the polarity of the applied RF signal and aperture length ($L$). The tips of the arrows indicate the thin electrodes on the ridge sidewalls. Scale bar 10 μm. **b** False-color SEM micrograph showing close-up and cross-sectional views of the resonating ridges. Scale bar 500 nm. **c** Measured admittance and $Q_{Bode}$ for VBARs with varying $w_{ridge}$ ($t_{oxide} = 0.5 \cdot t_{LNO}$ = 200 nm). **d** Comparison of measured admittance of three VBARs with different $SiO_2$ thicknesses, showing the effect of that layer on the magnitude of a spurious mode relatively close to anti-resonance. **e-g** Extracted **e** resonance frequency, **f** $k_t^2$ and **g** maximum $Q_{Bode}$ ($Q_{max}$) of the measured VBARs with $L = 70 \cdot w_{ridge}$. Markers represent mean values of 3 identical resonators. The errorbars represent the smallest and largest values of $Q_{max}$. **h** Comparison of measured admittance of a VBAR ($w_{ridge}$ = 600 nm) with a narrow anchor (as shown in **b**) and in direct-mount configuration.

Fig. 2a shows SEM micrographs of a fabricated one-port VBAR. We define the resonating ridges by patterning an array of large openings separated by the desired $w_{ridge}$ into the substrate. The etching process used for the ridges results in vertical sidewalls (85° to 90°, see Fig. 2b and Fig. S7 in the Supplementary Information). Next, we deposit a layer of Al conformally on the patterned substrate. With another lithography step, we cover the regions forming the pads and subsequently remove the Al on all exposed planar surfaces with dry etching. Following this self-aligned process, a thin film of Al remains on the sidewalls of the previously patterned holes, forming a ring that is connected on one edge to the pads. Hence, the structures between the holes form an array of parallel-connected VBAR ridges. We note that with the connection pattern shown in Fig. 2a, adjacent ridges are always driven with opposite phases (see also mode shapes in Fig. 1b). The fabrication is finalized by etching the partial undercut of the VBAR ridges. In Fig. 2b, we show a close-up view of two ridges of the fabricated VBAR, and a cross-sectional view obtained after cutting a ridge with Focused Ion Beam (FIB) milling. The cross-section view shows the shape of the Si anchor which is substantially narrower than the $LiNbO_3$ ridge. The amount of undercut is the same for all VBARs on the same substrate and fixed by the duration of the etching process. This means that the Si anchor is respectively wider or narrower relative to the width of the ridge for lower or higher frequency devices. The Al electrodes have a width of approximately 70 nm and widen slightly in the lower half of the ridge. In Fig. 2c, we show the measured admittance ($Y_{11}$) of VBARs with varying ridge widths, yielding resonance frequencies in the 2.5 to 4 GHz range with distances between $f_r$ and the anti-resonance frequency ($f_{ar}$) on the order of 500 MHz. We observe that the measured $Q$s (that is, Bode $Q$, $Q_{Bode}$, see also in the Methods) are generally below 150 and become gradually worse for narrower ridges and higher frequencies. Since the relative width of the anchors is largest for lowest $f_r$, this implies that leakage to the substrate is not a dominating factor in the observed $Q$s. Above $f_{ar}$, a prominent spurious peak arises from a higher order SH-BAW mode with thickness dependence (see Supplementary Information, Section 2). This mode can be suppressed by adjusting the $SiO_2$ thickness (see Fig. 2d). Between $f_r$ and $f_{ar}$, we observe evidence of transversal modes in the form of closely spaced ripples (see Fig. S4 in the Supplementary Information). In Fig. 2e-g, we summarize the extracted $f_r$, $k_t^2$, and $Q_{max}$. Qualitatively, the observed $f_r$ and $k_t^2$ match with the idealized simulations shown in Fig. 1. Non-uniformities in the thickness of the layers involved and other deviations from the initially conceived geometry (such as the sidewall angle and vertical position of electrodes) factor into the apparent discrepancies. In contrast to simulations, we observe a decrease in $f_r$ when removing the SiO$_2$ layer. Since Al is not etched during the SiO$_2$ removal process, the electrodes extend above the LiNbO$_3$ ridge, which lowers the frequency compared to the simulated case (flush electrode and ridge top surface, see Fig. S8 in the Supplementary Information). With $k_t^2$ of 30% or more, VBARs could potentially be useful for wide-bandwidth filters in the 5G mid-range. Even so, the quality factors we measure (Fig. 2g) are a major roadblock for a filter with low insertion loss. VBARs with the widest ridges (and as such, the lowest $f_r$) and largest total thickness exhibit the highest $Q$s. This trend suggests that the high losses could originate from surface loss mechanisms which are a key factor determining $Q$ in thin film LiNbO$_3$ resonators[41] and nanoelectromechanical resonators in general[42]. This explanation also seems probable considering that the sidewalls are visibly rough. Further, the cross-section of the ridges is close to a square which results in a significantly higher surface-to-volume ratio than a suspended film or a much taller ridge with the same width. At $f_r$, the $Q$ is further decreased by the resistive losses of the electrodes which represent roughly half of the impedance seen at resonance. To further demonstrate the energy confinement mechanism that the anchors provide, we fabricate a set of VBARs in which the ridge is directly mounted to the substrate without the anchors (Fig. 2h). In the directly mounted configuration, the peak admittance is significantly lower compared to the standard VBAR with an anchor and $Q_{\text{Bode}}$ never exceeds 30, which indicates poor energy confinement. This supports the notion that radiation to the bulk is reduced by the narrow anchors and that this loss mechanism is not the predominant reason for the overall low $Q$s.

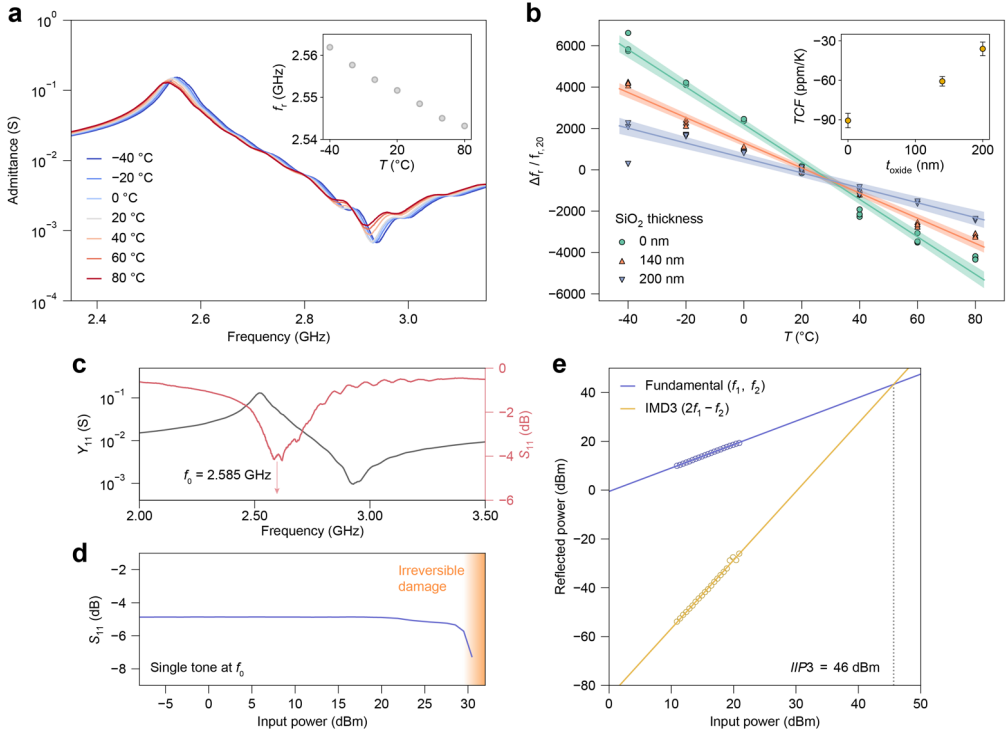

**Fig. 3. Measured temperature coefficient of frequency (*TCF*) and behavior at high power. a** Admittance response and extracted $f_r$ for varying temperature (*T*). **b** Calculated relative shift of $f_r$ with respect to $f_r$ at 20°C for VBARs with $w_{ridge}$ = 600 nm and varying SiO$_2$ thicknesses. For every *T*, we measure each resonator 3 times to minimize the influence of the contact on the TCF extraction. We perform linear fits to extract the first-order *TCF* from the slope of the fitted line. The shaded areas and error bars in the inset plot represent 95% confidence intervals of the fit. **c** Admittance ($Y_{11}$) and reflection coefficient ($S_{11}$) measured at low power (-15 dBm) of a VBAR ($w_{ridge}$ = 600 nm, $t_{oxide}$ = 0.35·$t_{LNO}$) before applying a high-power tone at $f_0$. **d** $S_{11}$ (of the resonator and probe) at $f_0$ for increasing input power. Starting at 29.5 dBm, we observe irreversible damage to the resonator. **e** IMD3 measurement of a VBAR ($w_{ridge}$ = 600 nm, $t_{oxide}$ = 0.35·$t_{LNO}$) with two input tones at $f_1$ = 2.617 GHz and $f_2 = f_1 + 2$ MHz. The linear extrapolations of the reflected fundamental and IMD3 power indicate a IIP3 value of 46 dBm.

In Fig. 3a, we show the admittance of a VBAR measured at temperatures ranging from -40 to 80°C. The measurements are performed in vacuum, but we do not see any notable increase in *Q* compared to atmospheric pressure. In Fig. 3b, we show the relative change in resonance frequency with respect to room temperature for VBARs with different thicknesses of SiO$_2$. The inclusion of SiO$_2$ lowers the TCF from -90 ppm/K to -36 ppm/K which corresponds in relative terms to the improvement predicted in simulation (-68 ppm/K to -27 ppm/K). Next, we assess the power handling capabilities of the VBARs. In Fig. 3c, we show the reflection coefficient and corresponding admittance of a VBAR measured at low power (-15 dBm) before applying high power signals. The return loss (RL) is at least 4 dB for a narrow band just above resonance, which indicates that the resonator under test absorbs a significant portion of the incoming power at those frequencies due to the rather low $Q$s. Using an in-house setup (described in Section 5 of the Supplementary Information), we apply a high-power tone at $f_0$ where the RL is maximized and measure the reflected power. Fig. 3d shows the measured $S_{11}$ of the resonator and probe as a function of incident power. Up to 29.5 dBm, $S_{11}$ is within 1 dB of the value measured at low power. Above 30 dBm, the resonators are irreversibly damaged and are eventually destroyed completely. In addition, using a two-tone setup, we measure the third-order intermodulation distortion (IMD3) induced by non-linear effects in the resonator (Fig. 3e). With the two tones centered at 2.618 GHz, we obtain an IIP3 value of 46 dBm. Further details on the power handling and IMD3 measurements are available in the Supplementary Information, Section 5.

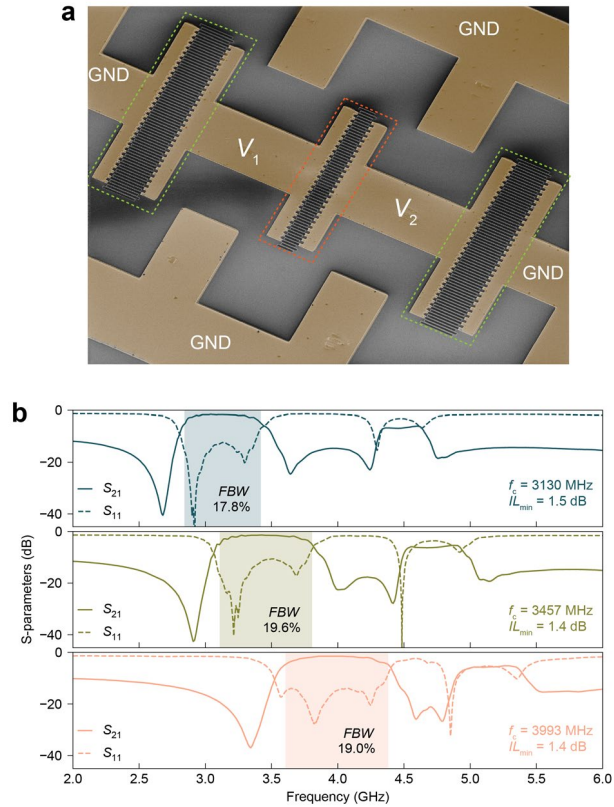

**Fig. 4. Bandpass filters using VBAR building blocks. a** False-color SEM images of a fabricated filter consisting of one series (red frame) and two shunt (green frame) VBARs. **b** Measured filter responses for three filters with different center frequencies ($f_c$). $f_c$ is set by varying $w_{ridge}$ lithographically for series and shunt resonators.

**VBAR filters**

We implement bandpass filters with VBAR impedance elements to demonstrate their potential for a wide-bandwidth and multi-frequency filtering application. As shown in Fig. 4a, we adopt a π-type filter topology containing two identical shunt resonators and one series resonator. We lithographically adjust $w_{\text{ridge}}$ of the VBARs to align $f_{\text{ar}}$ of the shunt to $f_{\text{r}}$ of the series resonator and control the center frequency of the filters. Crucially, we do not require any trimming of the film thicknesses to achieve the required frequency shifts. In Fig. 4b, we show the measured reflection and transmission coefficients ($S_{11}$ and $S_{21}$) for three filters with center frequencies ranging from 3.1 to 4 GHz, which we fabricate on the same chip. Due to the high $k_t^2$ of the VBARs, the FBWs of passband due to the SH-BAW resonances can reach nearly 20%. The spurious passband on the right of the intended passband is due to the presence of a higher order SH-BAW mode in the resonator response, which can be suppressed with a different choice of $SiO_2$ thickness (see also Fig. 2d and Fig. S5 in the Supplementary Information). The filters shown in Fig. 4b have minimum insertion losses of only 1 to 2 dB despite the relatively high overall losses in the resonators. However, this is mainly enabled by the choice of topology with only one series resonator which in turn only provides little out-of-band rejection. Improvements in $Q$ could drastically improve the flatness and roll-off slope of the passband and enable low insertion loss also in a higher-order filter.

**Discussion**

We have reported on an acoustic resonator architecture that attempts to combine the respective advantages of suspended and solidly mounted resonator configurations. We have shown that VBARs can deliver levels of $k_t^2$ (> 30%) that are comparable with fully suspended shear acoustic wave resonators such as XBARs or $LiNbO_3$ FBARs, while arguably providing better mechanical robustness due to the solidly mounted configuration of the resonator. Enabled by the ability to achieve large relative shifts in $f_{\text{r}}$ by lithographic design control, we have shown that VBARs are suitable for the implementation of filters with FBWs close to 20% without increasing the complexity of the fabrication.

Further, we have shown that VBARs can operate at power levels close to 30 dBm. While this is still considerably less than what commercial SAW[43] and AlScN-based BAW[21,44,45] resonators can achieve, it is a promising first result given that that the losses in the demonstrated prototypes are relatively high. In general, the acoustic losses in the resonator are the most significant roadblock that must be overcome before VBARs could be considered as fully suitable for more complex filter topologies. With the $Q$s that we have demonstrated, it would be challenging to design filters that comply with all the requirements for mid-range 5G bands. It is probable that the power handling could be improved if the losses and thus heat generation in the resonators are minimized. Further optimization of the fabrication process could reduce acoustic losses by lowering the surface roughness of the LiNbO$_3$ and electrode sidewalls. From a design perspective, using thicker LiNbO$_3$ films on the order of a micron or more may be beneficial for VBARs. The lower surface-to-volume ratio may lead to lower losses and the influence of any non-uniformities in the anchor dimensions on the resonance frequency would be reduced.

Furthermore, the general idea of a vertical, non-planar BAW resonator implies that the static capacitance of the devices could be decorrelated from the planar footprint of the resonator. Unlike FBARs or XBARs, the capacitance of such a resonator could be scaled by varying the thickness of the constitutive layers without significantly affecting the resonance frequency. This provides an additional degree of flexibility, enabling denser integration by reducing filter size or distributing heat over a larger substrate area under high power, depending on the specific application requirements.

## Methods

**Thin film LiNbO$_3$ on silicon substrates**

The substrates used for fabrication consist of a thin film of LiNbO$_3$ with a nominal thickness of 400 nm bonded to a high resistive silicon wafer purchased from NGK Insulators.

**Fabrication process flow**

Fig. S6 in the Supplementary information schematically outlines the process flow. As a first step, we deposit 1.7 microns of SiO$_2$ using plasma-enhanced chemical vapor deposition (PECVD, Corial D250L). We pattern the deposited SiO$_2$ layer using e-beam lithography followed by an inductively coupled plasma (ICP) based dry etching using a C$_4$F$_8$/H$_2$/He chemistry (SPTS Advanced Plasma

System). The patterned SiO$_2$ layer serves as a hard mask for the subsequent LiNbO$_3$ etching step. We etch the exposed LiNbO$_3$ using ion beam etching (IBE, Veeco Nexus IBE350) with a fixture angle of 45° which results in vertical sidewalls for the LiNbO$_3$ ridges (see Fig. S7 in the Supplementary Information). Next, we deposit and pattern a 300 nm Al layer serving as pads or filter interconnects with optical lithography and lift-off. To form the electrodes on the sidewalls, we first conformally deposit a 70 nm thick Al layer with sputtering. With another lithography step, we cover the regions of the pads and bus lines with photoresist. Subsequently, we use IBE with a fixture angle resulting in nearly vertical ion incidence (10°) to remove the previously sputtered Al in the exposed areas. This step removes Al only on horizontal surfaces (i.e. on top and next to the previously patterned LiNbO$_3$ ridges). The Al film that covers the vertical sidewalls of the ridges remains even though they are not masked. The next processing step consists of creating the Si pedestal that anchors the LiNbO$_3$ ridges to the substrate. We achieve this with an ICP-based dry etching step with SF$_6$ and a low frequency wafer voltage bias (Alcatel AMS 200 SE, see Fig. S7 in the Supplementary Information). For samples without a SiO$_2$ layer, the final processing step is the removal of the remaining SiO$_2$ covering the LiNbO$_3$ ridges with a selective dry etch with a C$_4$F$_8$/H$_2$/He chemistry (Fig. S8 in the Supplementary Information).

**S-parameters and admittance measurements**

All low power RF measurements are carried out with a Rohde & Schwarz ZNB20 Vector Network Analyzer (VNA) with a power of -15 dBm. For measurements at room temperature, MPI Titan T26 GSG probes with 200 microns pitch are used. For the variable temperature measurements, we use a LakeShore CRX-4K cryogenic probe station with cryo-compatible Picoprobe GSG probes (GGB Industries). Prior to all measurements, we calibrate the setup using a calibration substrate and a standard open-short-match (OSM) scheme. From the measured data, we compute the performance parameters of the resonator with $k_t^2 = \pi/2 \cdot f_r/f_{ar} \cdot \cot(\pi/2 \cdot f_r/f_{ar})$ and $Q_{Bode} = \omega \cdot |dS_{11}/d\omega| \cdot 1/(1-|S_{11}|^2)$[46]. The setups for the high power and non-linearity measurements are described in the Supplementary Information, Section 5.

**Data availability**

The data that support the findings of this study are available from the corresponding author upon reasonable request.

**Acknowledgments**

S.S. and L.G.V. thank V. Plessky and N. Zhang for helpful discussions. S. S. and L.G.V. acknowledge support from the Swiss National Science Foundation (SNSF) under projects CRSII5_189967 and 200020_184935. S.S. thanks the staff of the EPFL Center of MicroNanoTechnology (CMi) for their support with the fabrication process and tool maintenance.

**Ethics declarations**

**Conflict of Interest**

The authors declare no competing interest.

**Author contributions**

S.S. designed and conceptualized the implementation of the resonators and filters, carried out numerical simulations, fabricated the devices, conducted and analyzed the measurements, and wrote the manuscript. E.N., C.C. and J.M. performed and analyzed the non-linearity measurements. L.G.V supervised the work and acquired funding. All authors reviewed the draft of the manuscript.

# Supplementary Information:

# A bulk acoustic resonator with vertical electrodes for wideband filters

Silvan Stettler[1]*, Edgar Navarro-Gesse[2], Carlos Collado[2], Jordi Mateu[2], Luis G. Villanueva[1]

[1] Advanced Nanoelectromechanical Systems Laboratory, École Polytechnique Fédérale de Lausanne (EPFL), 1015 Lausanne, Switzerland

[2] Signal Theory and Communications Department, Universitat Politècnica de Catalunya (UPC), Castelldefels, Spain

* Corresponding author; silvan.stettler@epfl.ch

# Section 1: Suppression of acoustic energy leakage to the substrate

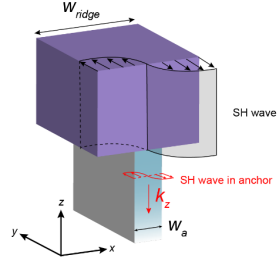

**Fig. S1 Simplified schematic representation of the resonator geometry.** The annotations show the displacements in the LiNbO$_3$ ridge (purple) and the Si pedestal acting as the anchor (light blue/gray). We assume the displacement to be a purely shear horizontal (SH) standing wave.

The general idea to suppress the leakage of acoustic energy is to create an anchor for the ridge in the form of a Si pedestal that does not support the transmission of the SH wave to the substrate. We assume that the resonating mode is a pure SH wave (displacement only along $z$, parallel to the boundaries of the ridge). Further, we simplify the anchor geometry and assume a narrow ridge with width $w_a$ (Fig. S1). In the absence of any mode conversion, the acoustic energy can only leak through the anchor as SH-type waves with a real wave vector component along $z$ ($k_z$). The cut-off frequency of the n-th order SH mode ($f_c^n$) in a rectangular waveguide with width $w_a$ is[1]

$$f_c^n = \frac{n \cdot v_{s,Si}}{2 \cdot w_a}$$

where $v_{s,Si}$ is the SH bulk wave velocity of Si (around 4600 m/s). Since the node of the displacement of the resonating mode coincides with the center line of the anchor, any transmitted waves in the anchor must be odd (*n* odd). The resonance frequency of the SH-wave in the ridge is approximately

$$f_r = \frac{v_{s,LNO}}{2 \cdot w_{ridge}}$$

where $v_{s,LNO}$ is the SH bulk wave velocity of LiNbO$_3$. To confine the SH wave in the ridge, the resonance frequency must be below the lowest cut-off frequency of the SH wave in the anchor which results in the condition

$$w_a < \frac{v_{s,Si}}{v_{s,LNO}} w_{ridge}$$

$v_{s,LNO}$ is smaller (3490 m/s and 4200 m/s for short and open-circuit boundary conditions, respectively) than $v_{s,Si}$. Thus, the resonance frequency is below the cut-off frequency of the anchor even if $w_a$ is equal to $w_{ridge}$, and the SH wave in the anchor is evanescent. Provided that the anchor is long enough to avoid evanescent coupling to BAWs in the substrate, the acoustic energy should remain confined in the LiNbO$_3$ ridge. It must be noted that this model is only approximative since it completely ignores any displacements associated with in-plane (vertical) shear or longitudinal partial waves.

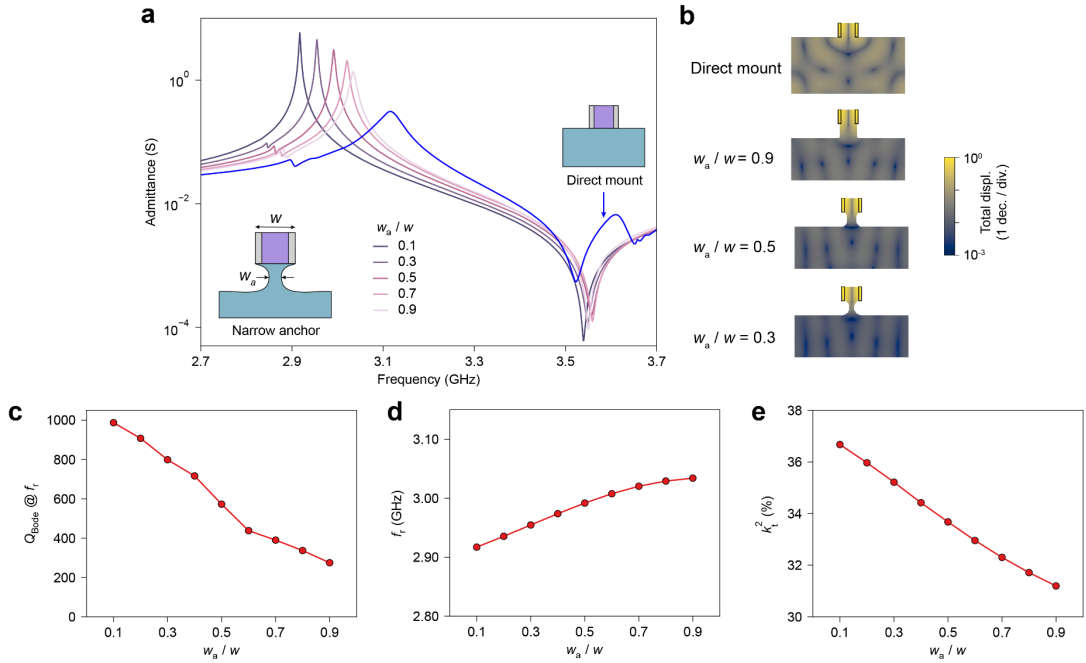

**Fig. S2 Simulated admittance and eigenmodes for different anchor dimensions.** Simulated admittance for the LiNbO$_3$ ridge **a** mounted directly to the silicon substrate (no anchor) and **b** mounted with an anchor with a width $w_a$ at the narrowest point. **c** Eigenmodes corresponding to the admittance peak for different anchor configurations. The displacement is plotted with a logarithmic normalization to emphasize the presence of BAWs in the substrate. **d-f** Simulated **d** resonance frequency ($f_r$), **e** $k_t^2$, and **f** $Q$ at resonance as a function of $w_a/w$. For the simulations in this figure, the other dimensions of the resonators are $w_{ridge}$ = 500 nm, $w_e$ = 100 nm, $t_{LNO}$ = 400 nm, $t_{oxide}$ = 100 nm. $w$ denotes the total width of the acoustic cavity (LiNbO$_3$ ridge plus twice the electrode width, $w_{ridge} + 2 \cdot w_e$). The intrinsic mechanical $Q$ assigned to all materials (imaginary part of stiffness coefficients) is 1000.

In Fig. S2a, we show the simulated admittance of a VBAR unit cell with the ridge mounted directly to the substrate and compared to a configuration with a narrow anchor of various relative widths. With the direct mount configuration, the ratio of admittances at resonance and anti-resonance is notably worse than with the narrow anchor configuration. In this case, the energy is only poorly confined, and the resonating ridge acts more like a transducer that radiates acoustic energy into the bulk. The admittance at resonance increases with decreasing anchor width. Simulations of the displacement field (Fig. S2b) show that the displacement amplitude in the bulk, which represents the leakage, is indeed decreasing the narrower the anchor is. In Fig. S2c, we show the $Q$ at resonance as a function of the relative anchor width. For $w_a/w = 0.1$, the $Q$ approaches the intrinsic $Q$ assigned to all materials in the simulation (= 1000). As shown in Fig. S1d-e, the dimensions of the anchor also affect $f_r$ and $k_t^2$. With a narrow anchor, the movement of the LiNbO$_3$ is less constrained, which results in higher $k_t^2$ and lower $f_r$.

# Section 2: Dependence of the VBAR response on the thickness

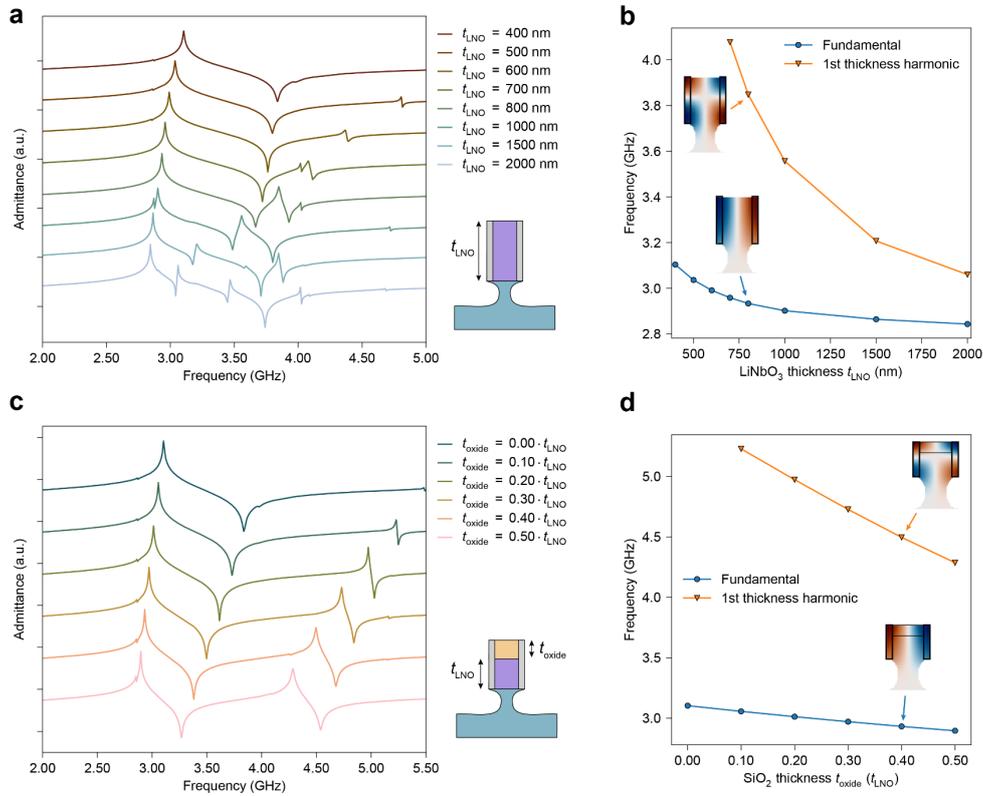

**Fig. S3 Simulations showing the dependence of the admittance response on the layer thicknesses. a** Admittance for varying LiNbO$_3$ thicknesses (without a SiO$_2$) layer. As $t_{LNO}$ increases, spurious thickness harmonics move closer to the fundamental resonance. **b** Resonance frequencies of the fundamental and the first thickness harmonic. **c** Admittance for a fixed $t_{LNO}$ = 400 nm and varying SiO$_2$ thickness ($t_{oxide}$). In contrast to the case without SiO$_2$ shown in **a-b**, the first thickness harmonic (ripple above 4 GHz) is stronger. **d** Resonance frequencies of the fundamental and first thickness harmonic for varying $t_{oxide}$. For these simulations, $w_{ridge}$ = 500 nm and $w_e$ = 100 nm.

## Section 3: Additional resonator and filter measurements

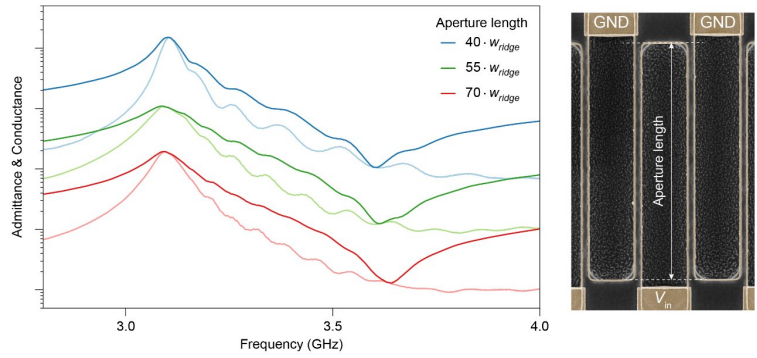

**Fig. S4 Measurements of VBARs with varying aperture lengths.** The measurements are shifted on the vertical axis for clarity. The spacing between the ripples appearing between $f_r$ and $f_{ar}$ becomes larger for shorter apertures which confirms that those ripples are transversal modes.

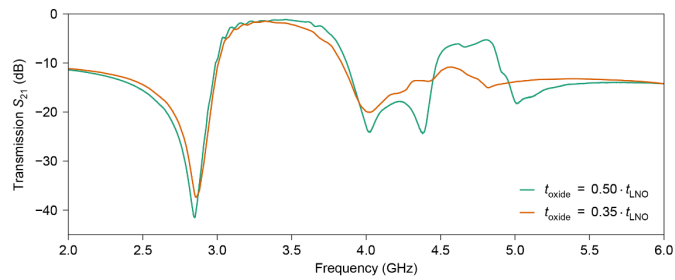

**Fig. S5 Filter response for different SiO$_2$ thicknesses.** For $t_{oxide} = 0.35 \cdot t_{LNO}$, the strength of the 1$^{st}$ thickness harmonic in the resonator response is reduced (Fig. 2d of the main text). This suppresses the formation of a spurious passband between 4.5 and 5 GHz in the filter response.

## Section 4: VBAR fabrication

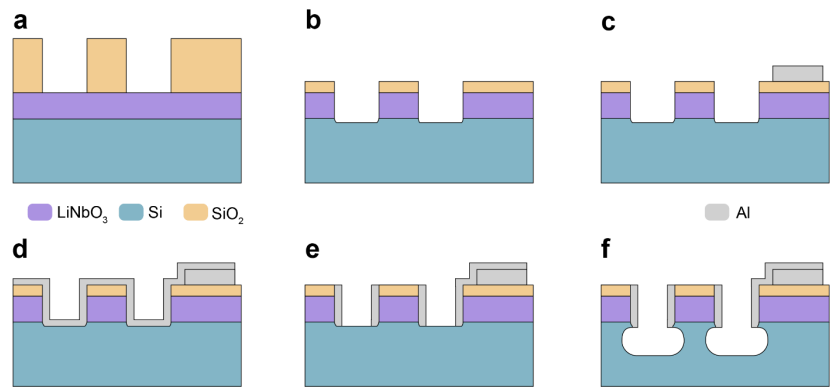

**Fig. S6 Schematic outline of the process flow used to fabricate the resonators and filters. a** SiO$_2$ deposition and patterning using e-beam lithography and ICP-based dry etching (CHF$_3$/H$_2$/He chemistry). **b** IBE with a 45° ion incidence angle for vertical sidewalls to define the LiNbO$_3$ ridges. **c** Optical lithography and lift-off for the Al pads. **d** Al sputtering followed by **e** IBE with near-vertical ion incidence (10°) to remove the Al only on horizontal surfaces. **f** Partial undercut of the ridges with dry etching (ICP SF$_6$ with a low frequency wafer voltage bias) to define the Si pedestals.

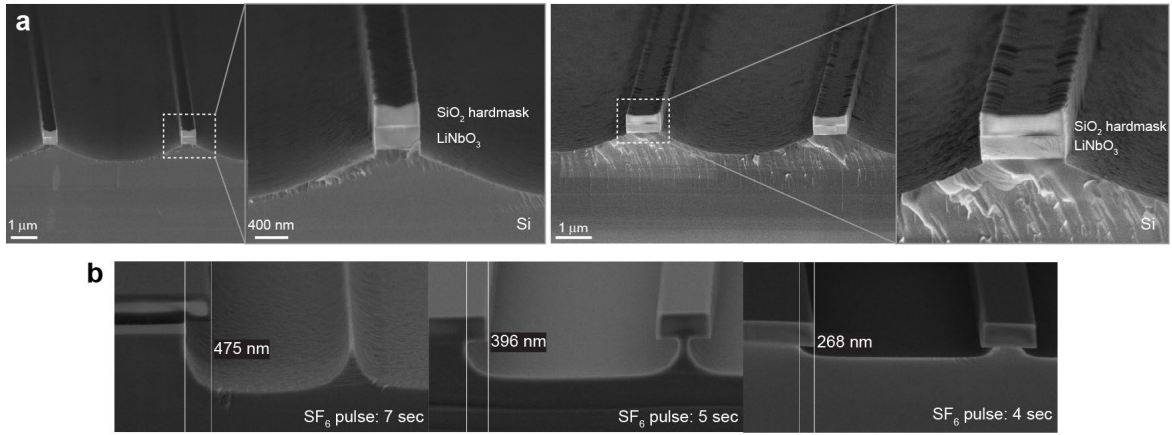

**Fig. S7 Details on the LiNbO$_3$ and silicon etching processes. a** Cleaved test sample after IBE of a thin LiNbO$_3$ film with the wafer fixture set for an ion incidence angle of 45° and a SiO$_2$ hard mask. With this fixture angle, the etch rate is significantly faster in the center of large features, but the sidewalls are vertical. **b** Test samples (patterned SiO$_2$ on Si wafer) showing how the width of the Si pedestal under the ridge can be controlled with the etch duration. The ICP-based recipe we use is designed for deep anisotropic etching of silicon but features a consistent undercut that correlates with the duration of the etch pulse. If the pulse is too long (7 sec.), the undercut is too large, and the ridge is fully released.

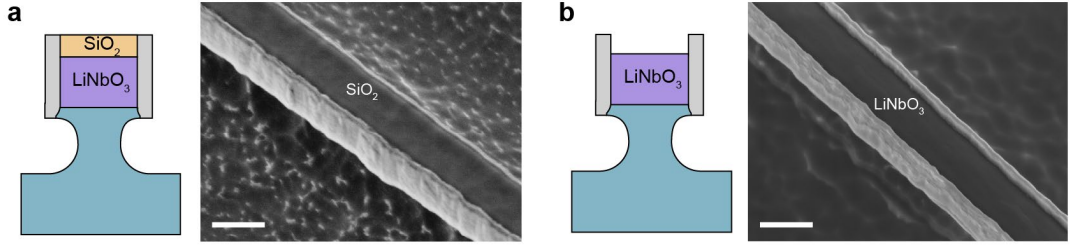

**Fig. S8 Ridge geometry before and after removing the SiO₂ layer.** SEM images and approximate cross-sectional geometries of the ridge **a** with the SiO₂ layer on top of the LiNbO₃ and **b** after removing the SiO₂ layer. The electrodes protrude above the LiNbO₃ top surface (instead of being flush) because they are not etched significantly by the dry etching process used to remove the SiO₂. Scale bar 500 nm.

## Section 5: Power handling and IMD3 measurements

Fig. S9a shows a block diagram of the 3-port setup used for power handling measurements. The input tone at $f_0$ from the signal generator (SG) passes through a high power amplifier and is transmitted to port 2 where a GSG probe and the device under test (DUT) are connected. The two circulators between the amplifier and the GSG probe isolate the amplifier output from the reflections of the DUT. The reflected signal at the DUT is transmitted towards a last circulator to attenuate and protect the spectrum analyzer (SA). Prior to measurements, we measure 3-port S-parameters (with the ports as defined on Fig. S9a) of the setup with a vector network analyzer (VNA). We assume that the SG and SA ports are perfectly matched. Using a signal flow graph, we relate the input power set with the signal generator ($P_{SG}$) to the outgoing power measured with the spectrum analyzer ($P_{SA}$). The power received by the SA (port 3) depends on the S-parameters of the setup and the reflection coefficient $\Gamma$ at port 2 (GSG probe plus the DUT):

$$P_{SA} = |S_{31} + S_{21}S_{32}\Gamma/(1-S_{22}\Gamma)|^2 \cdot P_{SG}$$

We solve this equation numerically to find $\Gamma$. $\Gamma$ represents the reflection coefficient of the DUT ($= S_{11}$) plus a small influence of the GSG probe, though the probe's good match and low insertion loss make this effect minimal. The incident power to the GSG probe is $P_{in} = |S_{21}|^2 P_{SG}$.

Fig. S9b shows a block diagram of the 4-port setup used for the IMD3 measurements. Prior to measurements, we measure 4-port S-parameters (with the ports as defined on Fig. S9b) of the setup with a VNA. The two input tones $f_1$ and $f_2$ (with power $P_{SG}$ each) are amplified and pass through a low-pass filter to remove the harmonics generated in the amplifiers. The two tones are combined with a Wilkinson combiner and transmitted towards the GSG probe and DUT with a hybrid coupler. The hybrid coupler directs the reflected power from port 3 to the SA. The incident power to the GSG probe (port 3) is $P_{in} = |S_{31}|^2 P_{SG}$. The power of the reflected fundamental and IMD3 tones at port 3 is calculated as $P_{refl}(f) = P_{SA}(f) / |S_{43}|^2$.

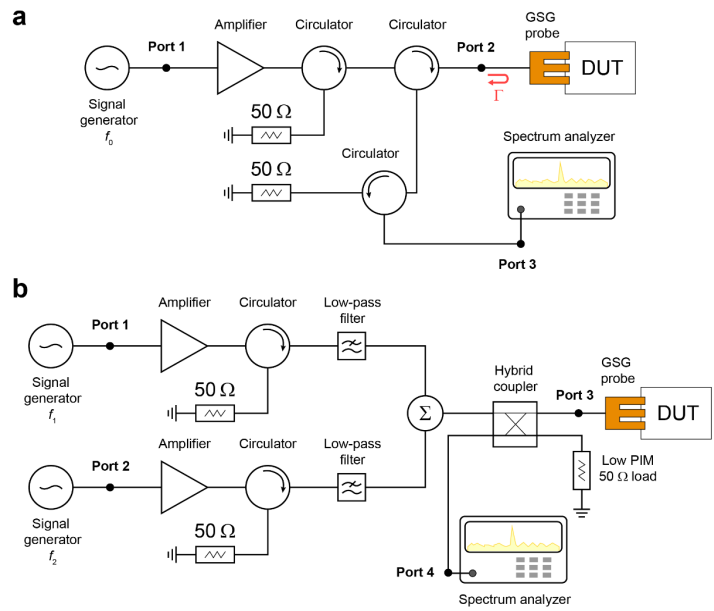

**Fig. S9 Experimental setups for high power and IMD3 measurements. a** 3-port setup with one input tone $f_0$ used to measure the reflection coefficient $\Gamma$ under high power. **b** 4-port setup with two input tones $f_1$ and $f_2$ used to measure IMD3.